# MULTILEVEL FUNCTIONAL PRINCIPAL COMPONENT ANALYSIS[1]

By Chong-Zhi Di, Ciprian M. Crainiceanu,
Brian S. Caffo and Naresh M. Punjabi

*Johns Hopkins University*

The Sleep Heart Health Study (SHHS) is a comprehensive landmark study of sleep and its impacts on health outcomes. A primary metric of the SHHS is the in-home polysomnogram, which includes two electroencephalographic (EEG) channels for each subject, at two visits. The volume and importance of this data presents enormous challenges for analysis. To address these challenges, we introduce multilevel functional principal component analysis (MFPCA), a novel statistical methodology designed to extract core intra- and inter-subject geometric components of multilevel functional data. Though motivated by the SHHS, the proposed methodology is generally applicable, with potential relevance to many modern scientific studies of hierarchical or longitudinal functional outcomes. Notably, using MFPCA, we identify and quantify associations between EEG activity during sleep and adverse cardiovascular outcomes.

**1. Introduction.**

1.1. *Data description.* The Sleep Heart Health Study (SHHS) is a large-scale comprehensive multi-site study of sleep and its correlation with health outcomes. In the following section we provide a detailed description of the study, and summarize some organizational and demographic characteristics. The principal aim of the study is to learn about the association between sleep and a variety of health-related conditions. The study is specifically designed to examine the potential associations between sleep-disordered breathing

Received May 2008; revised September 2008.
[1]Supported by Grants R01NS060910 from the National Institute of Neurological Disorders and Stroke, HL083640, HL07578, and AG025553 from the National Heart, Lung, and Blood Institute and K25EB003491 from the National Institute of Biomedical Imaging and BioEngineering.

(SDB) and outcomes such as hypertension and cardiovascular disease (CVD) [Quan et al. (1997)]. In our example analysis we focus on hypertension, a proposed consequence of disturbed sleep [Shahar et al. (2001)].

We now present a summary of the SHHS characteristics and our scientific hypotheses. A more detailed description of the SHHS can be found in Quan et al. (1997) and Crainiceanu et al. (2009). The SHHS is a multicenter study that utilized the resources of existing, well characterized, epidemiologic cohorts, and conducted further data collection, including measurements of sleep and breathing. These studies included the following: the Framingham Offspring and Omni Cohort Studies, the Atherosclerosis Risk in Communities Study (ARIC), the Cardiovascular Health Study (CHS), the Strong Heart Study, and the Tucson Epidemiologic Study of Respiratory Disease.

Between 1995 and 1997, a sample of 6,441 participants was recruited from the aforementioned parent studies. Participants less than 65 years of age were over-sampled on self-reported snoring to augment the prevalence of SDB. Prevalent cardiovascular disease (CVD) did not exclude potential participants and there was no upper age limit for enrollment. To acquire the sleep exposure variables, subjects underwent two in-home polysomnograms (PSGs), one at a baseline visit and one at a second visit, approximately five years later. A PSG is a quasi-continuous multi-channel recording of physiological signals acquired during sleep that include the following: two surface electroencephalograms (EEG), right and left electrooculograms for recording eye movements, leg and submentalis electromyograms, a precordial electrocardiogram, oxyhemoglobin saturation by pulse oximetry, and thoraco-abdominal movement with plethysmography. In addition to the in-home polysomnogram, extensive data on sleep habits, blood pressure, anthropometrics, medication use, daytime sleep tendency (ESS), and quality of life were collected. Outcome assessments were coordinated to provide standardized information on incident cardiovascular events.

*Baseline Characteristics of the SHHS cohort*: The baseline SHHS cohort of 6441 participants included 52.9% women and 47.1% men. Participants of Hispanic or Latino ethnicity comprised 4.5% of the sample. The race distribution in the sample was as follows: 81.0% Caucasians, 9.5% Native-Americans, 8.0% African-Americans, 1.3% Asians and 0.03% in the 'other race' category. The mean age of the cohort was 62.9 yr (SD: 11.0) and the mean body mass index (BMI) was 28.5 kg/m$^2$ (SD: 5.4). A modest number of participants were in the youngest (N = 750, age: 40–49 years) and oldest (N = 408, age > 80) age groups.

*Follow-up 1*: After the baseline visit, a follow-up examination of the cohort was conducted between 1998 and 1999 with assessment of vital status and other primary and secondary endpoints. Incident and recurrent cardiovascular events, medication use, sleep habits, blood pressure and anthropometry were assessed on surviving participants. The median follow-up time was 1.9 years with an interquartile range of 1.7 to 2.3 years.

*Follow-up 2*: A second SHHS follow-up visit was undertaken between 1999 and 2003 and included all of the measurements collected at the baseline visit along with a repeat PSG. The target population for the second follow-up exam included all surviving members who had a successful PSG at baseline. Exclusion criteria for the second PSG were similar to the criteria that were used at baseline, that is, conditions that pose technical difficulties for polysomnography. Although not all participants had a second PSG, 4361 of the surviving participants were recruited and completed the second SHHS visit (home visit with or without a PSG). A total of 3201 participants (47.8% of baseline cohort) completed a repeat home PSG. The median follow-up time was 5.2 years (interquartile range: 5.1–5.4 years).

With data on more than six thousand subjects at baseline, and over three thousand with repeat follow-up measurements, the size, complexity, and level of detail of these data are unprecedented in sleep research. For example, the study produced more than 1.5 terabytes of unprocessed EEG data, raising computational and methodological challenges. We further emphasize that there are a relatively limited number of published reports on EEG activity during sleep, most of which are based on smaller numbers of subjects (fewer than 50) and focus on clinical samples, rather than community based assessments [Sing et al. (2005); Tassi et al. (2006)]. Finally, there are only isolated studies using quantitative techniques to characterize EEGs during sleep as a function of age and gender, with the largest study consisting of only 100 subjects [Carrier et al. (2001)].

1.2. *Data processing and statistical challenges.* The first step of our analysis is to reduce the size of the data set by transforming it to the frequency space. This is necessary as the EEG data is sampled at a frequency of 125 Hz, resulting in $125 \times 60 \times 60 \times (\textit{hours of sleep})$ (e.g., 6–8 hours) data points for an eight hour sleep interval, per subject, per channel, per visit. The transformation to the frequency domain was also needed because specific frequency bands are of interest to sleep researchers.

The original quasi-continuous EEG signal was pre-processed using the Discrete Fourier Transform (DFT). More precisely, if $x_0, \ldots, x_{N-1}$ are the $N$ measurements from a raw EEG signal, then the DFT is $F_{x,k} = \sum_{n=0}^{N-1} x_n e^{-2\pi i n k / N}$ for $k = 0, \ldots, N-1$, where $i$ is the imaginary unit. If $R$ denotes a range of frequencies, then the power of the signal in the $R$ frequency range is defined as $P_R = \sum_{k \in R} F_{x,k}^2$. Four frequency bands were of interest: (1) $\delta$ [0.8–4.0 Hz]; (2) $\theta$ [4.1–8.0 Hz]; (3) $\alpha$ [8.1–13.0 Hz]; (4) $\beta$ [13.1–20.0 Hz]. These bands are standard representations of low ($\delta$) to high ($\beta$) frequency neuronal activity. For the current analysis, we focus on $\delta$ power. However, to make $\delta$ power comparable across subjects, we normalized it as $\mathrm{NP}_\delta = P_\delta/(P_\delta + P_\theta + P_\alpha + P_\beta)$. The normalized $\delta$-power is thought to be less dependent on the amplitude of the signal, which can be influenced by potential drift that may occur over the course of the night, particularly with unattended monitoring in the home setting. Because of the nonstationary nature of the EEG signal, the DFT and normalization were applied in adjacent 30-second intervals, resulting in the temporal representation: $t \to \mathrm{NP}_\delta(t)$, where $t$ indicates the mid-point of the corresponding 30-second interval.

To better understand the data structure, Figure 1 displays the fraction of $\delta$-power for three subjects at two visits in the SHHS. The dots represent pairs $\{t, \mathrm{NP}_\delta(t)\}$ and the solid lines represent the estimated mean function using penalized splines. These data raise a range of challenges that are characteristic of many modern data sets. First, data are functions that exhibit large within- and between-subject heterogeneity. Second, the underlying mean functions are clustered within subjects. Third, the signal is measured with sizeable error. Fourth, the dataset is very large, with two visits for more than 3000 subjects.

It is desired to use the $\delta$-power as a predictor of adverse outcomes, such as hypertension or CVD. However, it is a challenging topic to use the hierarchical functions appropriately as predictors. A viable alternative was employed by Crainiceanu et al. (2009), who used interpretable core features of smoothed versions of the subject-specific functions, such as the maximum and the time to maximum, as predictors. Here, we take a different approach and provide a complete framework for the analysis of multilevel functional data with application to the SHHS. Our methods, which are described in Section 2, are based on the decomposition of functional variability according to the natural hierarchy induced by the sampling mechanism. The practical goals of our analyses are to: (1) provide a robust and computationally efficient method for dimensionality reduction and signal extraction from multilevel functional data, (2) provide a geometric representation of multilevel functional data, (3) quantify the variability corresponding to each level of the hierarchy and residual noise, and (4) quantify the signals that could replace the functions without major loss of information in subsequent analyses.

1.3. *Methods for the analysis of functional data.* The statistical framework of functional data analysis (FDA) is a term introduced by Ramsay and Silverman. Their popular book [Ramsay and Silverman (2005)] provides a broad overview of functional data analysis methods with applications to curve and image analysis. The standard inferential methods for the analysis of functional data can be divided into two general areas: functional linear models (including functional analysis of variance, Functional ANOVA) and functional principal component analysis (FPCA).

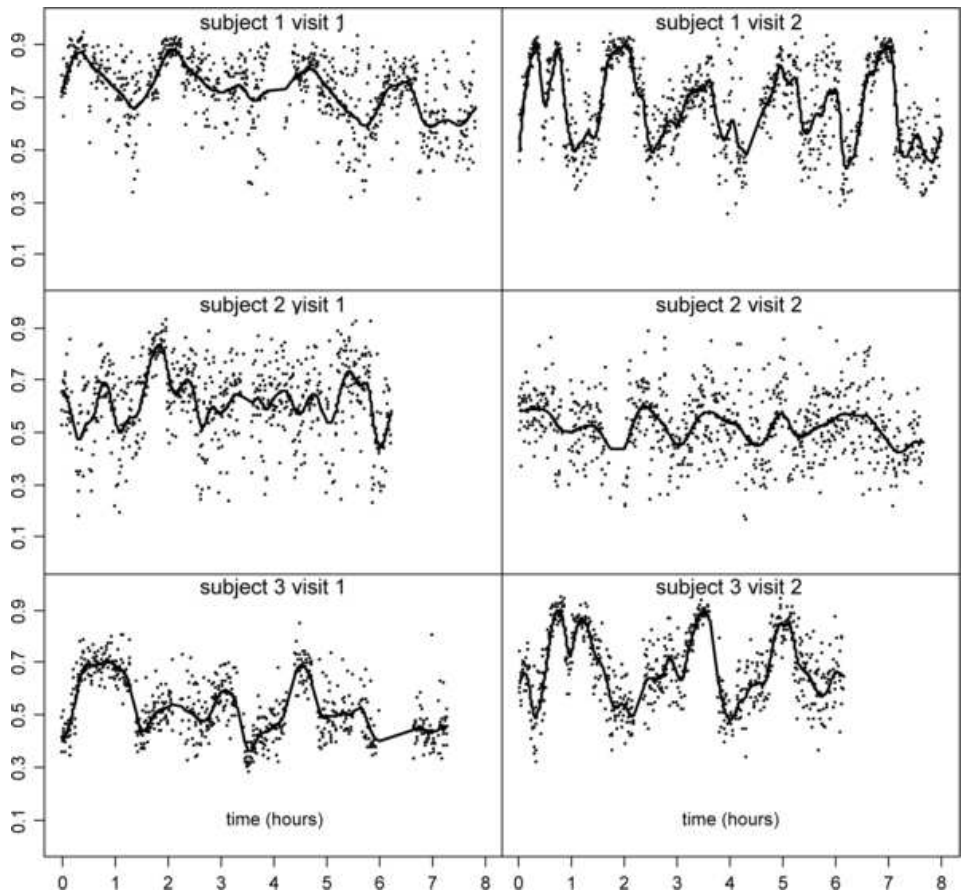

FIG. 1. *Normalized sleep EEG $\delta$-power for the first 3 subjects at 2 visits. X-axis: time in hours since sleep onset. Y-axis: percentage of sleep EEG $\delta$-power in 30-second windows. Solid lines: subject/visit specific penalized splines estimators of the mean signal. The full dataset contains more than 3000 such subjects.*

Functional linear models provide an elegant statistical framework for functional variability decomposition. In particular, Functional ANOVA models are designed for inference on level-specific functional means, given a natural hierarchy of units (e.g., subject-specific exposure markers within treatments). For example, smoothing spline models for nested and crossed curves were introduced in Brumback and Rice (1998). Functional mixed effects models were discussed in Guo (2002). Bayesian wavelet models were proposed in Morris and Carroll (2006) and Morris et al. (2003). Bayesian adaptive regression splines were introduced in Bigelow and Dunson (2007), and Bayesian models for spatially correlated functional data were analyzed in Baladandayuthapani et al. (2008). These and other important method-ological research efforts have resulted in a rich collection of inferential methods for functional linear models. However, these methods are not directly applicable to the SHHS for several reasons. First, functional ANOVA models are focused on estimating level-specific means for functional data. In contrast, for the SHHS we are interested in subject-specific signal extraction. Subject-specific functional signals are subsequently used in second-level analyses. Second, functional ANOVA uses clustering of curves around group averages, where the group membership is well defined, such as with treatment assignment in a clinical trial. In SHHS, group membership is not well defined, or may be the actual object of inference. Moreover, functional ANOVA models have been applied to studies with a smaller number of subjects. The only exception is the recent work on wavelet based functional mixed models Morris and Carroll (2006) and Morris et al. (2008). For example, Morris and Carroll (2006) contains an application with 750 functions on a grid of 256 and Morris et al. (2008) contains two applications, one of them with 32 curves each on a grid of about 8000, and the other with 256 curves each on a grid of about 12,000. Our methods have been applied to 6000 curves (3000 subjects at two visits) on a grid of about 480. In simulations we successfully implemented our analyses to data sets with 20,000 curves.

The second general area is functional principal component analysis (FPCA). The fundamental aims of this method include capturing the principal directions of variation and dimension reduction. FPCA summarizes the subject-specific features as the coordinates (called principal component scores) of subject curves in the basis spanned by the principal components. Besides discussion in Ramsay and Silverman (2005), other relevant research in FPCA includes Ramsay and Dalzell (1991), Silverman (1996), James, Hastie and Sugar (2000), and Yao, Müller and Wang (2005), while important theoretical results can be found in Hall and Hosseini-Nasab (2006).

An important limitation of current FPCA methods is that they are not designed for multilevel analyses, such as when subject-level curves are observed at several visits. There has been considerable recent effort to apply FDA to longitudinal data [see Müller (2005), for a thorough review]. These methods were developed for one or more functions that are sparsely observed over a single time course, such as height being observed over childhood in growth studies. Thus, the term "longitudinal" is used to refer to a single-level time series. In contrast, the data in the SHHS has two time distinct time courses, the time over a night of sleep and the time over the repeated collection of sleep data. That is, we consider longitudinally collected longitudinal functions. Except for our own preliminary studies, we are unaware of FPCA methodology for studying such datasets.

The remainder of the paper is organized as follows. Section 2 introduces Multilevel Functional Principal Component Analysis (MFPCA), our statistical framework for multilevel functional data (MFD). Section 3 provides the methodology for estimating the functional scores. Section 4 describes extensive simulation studies for realistic settings. Section 5 describes the application of our methodology to the SHHS data set. Section 6 presents our conclusions. To ensure reproducibility of our results accompanying software, simulations and analysis results are posted at http://www.biostat.jhsph.edu/˜ccrainic/webpage/software/MFD.zip.

**2. Models and framework.** We first review the widely used functional principal component analysis (FPCA) technique [as discussed in Ramsay and Silverman (2005)]. FPCA plays a central role in Functional Data Analysis (FDA) and is designed to describe the geometry of functions when one function is available per subject. The basic idea of FPCA is to decompose the space of curves into principal directions of variation.

Let $X(t)$, $t \in [0,1]$, be a squared integrable random function with mean $\mu(t)$ and covariance function $K(s,t)$; that is, $\mu(t) = E\{X(t)\}$ and $K(s,t) = \text{cov}\{X(s), X(t)\}$. Mercer's theorem [see Indritz (1963), Chapter 4] provides the following convenient spectral decomposition of $K(s,t)$:

$$K(s,t) = \sum_{k=1}^{\infty} \lambda_k \phi_k(s) \phi_k(t),$$

where $\lambda_1 \geq \lambda_2 \geq \cdots$ are ordered nonnegative eigenvalues and $\phi_k$'s are the corresponding orthogonal eigenfunctions with unit $L^2$ norms. Since the eigenfunctions form a basis for $L^2[0,1]$ functional space, the Karhunen–Loève (KL) expansion [Karhunen (1947); Loève (1945)] of the random function $X(t)$ is $X(t) = \mu(t) + \sum_{k=1}^{\infty} \xi_k \phi_k(t)$, where $\xi_k = \int_0^1 \{X(t) - \mu(t)\} \phi_k(t)\,dt$ are uncorrelated random variables with mean zero and variance $\lambda_k$. These random variables are called principal component scores or loadings. The KL expansion provides the theoretical platform for FPCA. For a given functional sample, the mean function, $\mu(t)$, and covariance function, $K(s,t)$, can be consistently estimated using, for instance, the method of moments. The eigenvalues and eigenfunctions are estimated from the empirical covariance function, and the principal component scores can be estimated by numerical integration. In practice, only a few eigenvalues and eigenfunctions are needed to capture the important modes of variations of a sample of random functions.

2.1. *Multilevel FPCA.* Recall that our primary interest lies in clustered or multilevel functional data. For example, consider the motivating SHHS application, where it is natural to assume that the normalized $\delta$-power functions at baseline and the second visit are correlated. To appropriately address this correlation, we propose Multilevel FPCA (MFPCA), a new framework evolved from a combination, of FPCA and standard multilevel mixed models, though presenting unique statistical challenges that are addressed below. We do not provide an overview of multilevel models here, but instead point readers to a few excellent monographs: Diggle et al. (2002), Goldstein (1995), and Raudenbush and Bryk (2002).

Notationally, let $X_{ij}(t)$ be a function measured over a continuous variable $t$ for observation $j$ within cluster $i$ for $i = 1, 2, \ldots, I$ and $j = 1, 2 \ldots, J$. In our application, $t$ is time from sleep onset, $i$ is subject, and $j$ is visit. Without loss of generality, we restrict attention to the case when each subject is measured for every value of $j$, but emphasize that our methodology does not require this assumption.

As a first step, consider the two-way functional ANOVA model

$$(2.1) \qquad X_{ij}(t) = \mu(t) + \eta_j(t) + Z_i(t) + W_{ij}(t),$$

where $\mu(t)$ is the overall mean function, $\eta_j(t)$ is the visit-specific shift from the overall mean function, $Z_i(t)$ is the subject-specific deviation from the visit-specific mean function, and $W_{ij}$ is the residual subject- and visit-specific deviation from the subject-specific mean. In the SHHS $\mu(t)$ and $\eta_j(t)$ are treated as fixed functions, while $Z_i(t)$ and $W_{ij}(t)$ are treated as mean 0 stochastic processes. The former is a reasonable assumption in our application, as the SHHS contains more than 3000 subjects who completed two visits. We also assume that $Z_i(t)$ and $W_{ij}(t)$ are uncorrelated. We note that, in many applications, $\eta_j(t)$ could be set to a zero when functional responses are exchangeable within clusters and the model becomes a one-way functional ANOVA; for example, when considering siblings in a family or patients in a clinic. As it is of interest in our application, $\eta_j(t)$ will be estimated.

This model is the "hierarchical functional model" introduced by Morris et al. (2003) and also used by Bigelow and Dunson (2007) and Baladandayuthapani et al. (2008). It is also a special case of the general "functional mixed model" used in Morris and Carroll (2006). Prior to these works, the functional ANOVA models did not include the possibility of multiple nested levels of random effect functions, which is what separates our work from other functional principal component literature.

Using the intuition from typical multilevel models, we call $Z_i(t)$ level 1 functions, and $W_{ij}(t)$ level 2 functions. The core idea of MFPCA is to decompose both level 1 and level 2 functions using the KL expansion. More precisely, let

$$(2.2) \qquad Z_i(t) = \sum_k \xi_{ik} \phi_k^{(1)}(t), \qquad W_{ij}(t) = \sum_l \zeta_{ijl} \phi_l^{(2)}(t),$$

where $\xi_{ik}$ and $\zeta_{ijl}$ are level 1 and level 2 principal component scores respectively, and $\phi_k^{(1)}(t)$ and $\phi_l^{(2)}(t)$ are level 1 and level 2 eigenfunctions, respectively. Model (2.1) with the KL expansions (2.2) becomes

$$(2.3) \quad X_{ij}(t) = \mu(t) + \eta_j(t) + \sum_{k=1}^{\infty} \xi_{ik} \phi_k^{(1)}(t) + \sum_{l=1}^{\infty} \zeta_{ijl} \phi_l^{(2)}(t),$$

where $\mu(t), \eta_j(t), \phi_k^{(1)}(t), \phi_k^{(2)}(t)$ are fixed functional effects, and the $\xi_{ik}$ and $\zeta_{ijl}$ are zero mean random variables. At a first glance, model (2.3) may appear too complex to be implemented for studies with large sample size, such as the SHHS. However, we show that inference from this model can be done using a short sequence of simple steps.

We summarize the core assumptions as follows:

(A.1) $E(\xi_{ik}) = 0$, $\mathrm{var}(\xi_{ik}) = \lambda_k^{(1)}$, for any $i$, $k_1 \neq k_2$, $E(\xi_{ik_1} \xi_{ik_2}) = 0$;
(A.2) $\{\phi_k^{(1)}(t) : k = 1, 2, \ldots\}$ is an orthonormal basis of $L^2[0,1]$;
(A.3) $E(\zeta_{ijl}) = 0$, $\mathrm{var}(\zeta_{ijl}) = \lambda_l^{(2)}$, for any $i$, $j$, $l_1 \neq l_2$, $E(\zeta_{ijl_1} \zeta_{ijl_2}) = 0$;
(A.4) $\{\phi_l^{(2)}(t) : l = 1, 2, \ldots\}$ is an orthonormal basis of $L^2[0,1]$;
(A.5) $\{\xi_{ik} : k = 1, 2, \ldots\}$ are uncorrelated with $\{\zeta_{ijl} : l = 1, 2, \ldots\}$.

Assumptions (A.1)–(A.4) are standard for functional principal component analysis, and (A.5) corresponds to the previously stated assumption that $Z_i(t)$ and $W_{ij}(t)$ are uncorrelated. Note that the level 1 and 2 eigenfunctions, $\{\phi_k^{(1)}(t) : k = 1, 2, \ldots\}$ and $\{\phi_l^{(2)}(t) : l = 1, 2, \ldots\}$, are assumed to be orthonormal bases, but are not required to be mutually orthogonal.

2.2. *Eigenvalues and eigenfunctions.* We now focus on estimating the eigenvalues and eigenfunctions in model (2.3) under assumptions (A.1)–(A.5). Let $K_T(s,t) = \mathrm{cov}\{X_{ij}(s), X_{ij}(t)\}$ be the overall covariance function, and $K_B(s,t) = \mathrm{cov}\{X_{ij}(s), X_{ik}(t)\}$ be the covariance function between level 2 units within the same level 1 unit. Then $K_T(s,t) = \sum_{k=1}^{\infty} \lambda_k^{(1)} \phi_k^{(1)}(s) \phi_k^{(1)}(t) + \sum_{l=1}^{\infty} \lambda_l^{(2)} \phi_l^{(2)}(s) \phi_l^{(2)}(t)$ and $K_B(s,t) = \sum_{k=1}^{\infty} \lambda_k^{(1)} \phi_k^{(1)}(s) \phi_k^{(1)}(t)$. Define $K_W(s,t) := K_T(s,t) - K_B(s,t)$, that is, $K_W(s,t) = \sum_{l=1}^{\infty} \lambda_l^{(2)} \phi_l^{(2)}(s) \phi_l^{(2)}(t)$, where the indices $T$, $B$, and $W$ are used to refer to the "total," "between," and "within" subject covariances, respectively. These, of course, are not the same quantities as in mixed ANOVA models, but our notation builds upon the intuitive variance decomposition of these simpler models. The within and between decomposition of variability based on the KL expansion leads to the following convenient algorithm:

*Step* 1 estimate the mean and covariance function, $\hat{\mu}(t), \hat{\eta}_j(t)$, $\hat{K}_T(s,t)$ and $\hat{K}_B(s,t)$ using the method of moments; set $\hat{K}_W(s,t) = \hat{K}_T(s,t) - \hat{K}_B(s,t)$;
*Step* 2 use eigenanalysis on $\hat{K}_B(s,t)$ to obtain $\hat{\lambda}_k^{(1)}$, $\hat{\phi}_k^{(1)}(t)$;
*Step* 3 use eigenanalysis on $\hat{K}_W(s,t)$ to obtain $\hat{\lambda}_l^{(2)}$, $\hat{\phi}_l^{(2)}(t)$;

*Step* 4 estimate principal component scores (technical details in Section 3).

In practice, each function, $X_{ij}(t)$, is measured at a set of grid points $\{t_{ijs} : s = 1, 2, \ldots, T_{ij}\}$. If these sampling points are common for every subject and visit, that is, $t_{ijs} = t_s$ and $T_{ij} = T$, then the method of moment estimators in *Step 1* of the algorithm is easy to construct. More precisely, $\hat{\mu}(t_s) = \bar{X}_{..}(t_s)$ and $\hat{\eta}_j(t_s) = \bar{X}_{\cdot j}(t_s) - \bar{X}_{..}(t_s)$, where $\bar{X}_{..}(t_s) = \sum_{i,j} X_{ij}(t_s)/(IJ)$ and $\bar{X}_{\cdot j}(t_s) = \sum_i X_{ij}(t_s)/I$. Furthermore, $\hat{K}_T(t_s, t_r) = \sum_{i,j} \{X_{ij}(t_s) - \hat{\mu}(t_s) - \hat{\eta}_j(t_s)\}\{X_{ij}(t_r) - \hat{\mu}(t_r) - \hat{\eta}_j(t_r)\}/(IJ)$ and $\hat{K}_B(t_s, t_r) = 2 \sum_i \sum_{j_1 < j_2} \{X_{ij_1}(t_s) - \hat{\mu}(t_s) - \hat{\eta}_{j_1}(t_s)\}\{X_{ij_2}(t_r) - \hat{\mu}(t_r) - \hat{\eta}_{j_2}(t_r)\}/\{IJ(J-1)\}$. The method of moments estimators can be constructed in a variety of other situations. If the sampling points are reasonably dense for each subject/visit, then data can be smoothed first and the mean predicted on an equally spaced grid of points.

A different case occurs when data for each subject/visit is sparse, but the collection of sampling points over subjects and visits is dense. A reasonable approach in this case would be to consider the histogram of all sampling points, $t_{ijs}$, using, for example, a fine grid of quantiles. Each time point can then be approximated by the time center of its corresponding bin. For the case of sparse data an alternative strategy could be using smoothing techniques. For instance, the overall mean function $\mu(t)$ could be estimated by smoothing the pairs $\{(t_{ijs}, X_{ij}(t_{ijs})) : i = 1, \ldots, I; j = 1, \ldots, J; s = 1, \ldots, T_{ij}\}$, and $\eta_j(t)$ could be estimated similarly. The covariance functions could be estimated by bivariate smoothers, for example, the between covariance function $K_B(t_{ij_1s}, t_{ij_2r})$ is estimated by smoothing $\{X_{ij_1}(t_{ij_1s}) - \hat{\mu}(t_{ij_1s}) - \hat{\eta}_{j_1}(t_{ij_1s})\}\{X_{ij_2}(t_{ij_2r}) - \hat{\mu}(t_{ij_2r}) - \hat{\eta}_{j_2}(t_{ij_2r})\}$ with respect to $(t_{ij_1s}, t_{ij_2r})$. This method is inspired by Yao, Müller and Wang (2005) for sparse single-level functional data, but will not be pursued in details here. In the SHHS, the data are equally spaced and completely observed.

Because $K_W(s, t)$ in *Step* 1 is estimated as a difference, it may not be positive definite. This problem can be solved by trimming eigenvalue-eigenvector pairs where the eigenvalue is negative [Hall, Müller and Yao (2008); Müller (2005), Yao, Müller and Wang (2005)]. As shown in Hall, Müller and Yao (2008), this method is more accurate than the method of moments in terms of the $L^2$ norm. In all our simulations in Section 4 and in the SHHS application in Section 5, the magnitude of the negative eigenvalues is very small relative to the positive eigenvalues, and the procedure performed remarkably well.

Choosing the number of eigenfunctions is an important practical problem without a theoretically satisfactory solution. Two practical alternatives are to use cross validation [Rice and Silverman (1991)] or Akaike's Information Criterion [or AIC, as done in Yao, Müller and Wang (2005)]. One might choose an even simpler method for estimating the number of components based on the estimated explained variance. More precisely, let $P_1$ and $P_2$ be two thresholds and define

$$N_1 = \min\{k : \rho_k^{(1)} \geq P_1, \lambda_k < P_2\},$$

where $\rho_k^{(1)} = (\lambda_1^{(1)} + \cdots + \lambda_k^{(1)})/(\lambda_1^{(1)} + \cdots + \lambda_T^{(1)})$. For the cumulative explained variance threshold, we used $P_1 = 0.9$ and for the individual explained variance, we used $P_2 = 1/T$, where $T$ is the number of grid points. We used a similar method for choosing the number of components at level 2. These choices were slightly conservative, but worked well in our simulations and application. However, the two thresholds should be carefully tuned in any other particular application using simulations.

An important parameter is the proportion of variability explained by level 1, which is the variance explained by the within cluster variability. From equation (2.2), the sum of eigenvalues at a particular level is the average variance of functions at that level. More precisely, $\sum_{k=1}^{\infty} \lambda_k^{(1)} = \int \text{var}\{Z_i(t)\} dt$ and $\sum_{l=1}^{\infty} \lambda_l^{(2)} = \int \text{var}\{W_{ij}(t)\} dt$. A natural measure of variance explained by within cluster variability is

$$(2.4) \quad \rho_W = \frac{\sum_{k=1}^{\infty} \lambda_k^{(1)}}{\sum_{k=1}^{\infty} \lambda_k^{(1)} + \sum_{l=1}^{\infty} \lambda_l^{(2)}} = \frac{\int \text{var}\{Z_i(t)\} dt}{\int \text{var}\{Z_i(t)\} dt + \int \text{var}\{W_{ij}(t)\} dt},$$

which is the functional analogue of the intra-cluster correlation in standard mixed effects ANOVA models.

2.3. *Smooth MFPCA for functions measured with error.* In the previous sections we assumed that the functions are perfectly observed. However, in many applications, including the SHHS, functional signals are measured with error. For standard FPCA, a survey of the literature reveals three methods for addressing this problem. The first approach is to smooth the data before applying FPCA [see Besse and Ramsay (1986); Ramsay and Dalzell (1991)]. The second is to introduce a penalty term for FPCA [Silverman (1996); Ramsay and Silverman (2005)]. The third is to smooth the covariance function [Yao et al. (2003); Yao, Müller and Wang (2005)].

Our approach is similar in spirit to the latter method, but with important differences, to take into account the additional complexity induced by functional clustering. Notationally, we assume that we observe noisy data $Y_{ij}(t) = X_{ij}(t) + \varepsilon_{ij}(t)$, where $X_{ij}(t)$ is assumed to come from model (2.3) and $\varepsilon_{ij}(t)$ is a white noise process with variance $\sigma_i^2$. For simplicity of presentation, we assume that $\sigma_i^2 = \sigma^2$ for all $i$, but emphasize that the methodology is not limited by this assumption. The covariance functions of $Y_{ij}(t)$'s, $G_T(t, s) = \text{cov}\{Y_{ij}(s), Y_{ij}(t)\}$ and $G_B(t, s) = \text{cov}\{Y_{ij}(s), Y_{ik}(t)\}$ are related to the covariance functions of the $X_{ij}(t)$'s through

$$G_T(t,s) = \text{cov}\{X_{ij}(s), X_{ij}(t)\} + \sigma^2 \text{I}(t=s)$$
$$= K_T(t,s) + \sigma^2 \text{I}(t=s); \tag{2.5}$$

$$G_B(t,s) = \text{cov}\{X_{ij}(s), X_{ik}(t)\} = K_B(t,s). \tag{2.6}$$

These equations suggest a simple solution for estimating the eigenvalues, eigenfunctions and the nugget variance, $\sigma^2$. The first step is to estimate the mean functions $\mu(t)$ and $\eta_j(t)$ using either local polynomial smoothing [Fan and Gijbels (1996)] or penalized spline smoothing [Ruppert et al. (2003)] under the working independence assumption. The choice of smoothing parameters is well discussed in the smoothing literature. For instance, cross validation is popular for the former, while REML (restricted maximum likelihood) or GCV (generalized cross validation) works well for the latter. For issues on smoothing for dependent data, see the discussion in Lin and Carroll (2000). The second step is to obtain the method of moment estimates of $G_T(t,s)$ and $G_B(t,s)$, denoted by $\hat{G}_T(t,s)$ and $\hat{G}_B(t,s)$, respectively. The third step is to estimate $\hat{K}_T(t,s)$ by smoothing $\hat{G}_T(t,s)$ for $t \neq s$. The idea of dropping diagonal elements in smoothing was from Staniswalis and Lee (1998) and Yao, Müller and Wang (2005). An inspection of equation (2.5) reveals why the diagonal elements $\hat{G}_T(t,t)$ should be removed in this step. $\hat{K}_B(t,s)$ is estimated by smoothing $\hat{G}_B(t,s)$ for all $t$ and $s$, according to equation (2.6). For both smoothing procedures we use penalized thin plate spline smoothing with the smoothing parameter estimated via REML. The fourth step is to predict the diagonal elements, $\hat{K}_T(t,t)$, and estimate the error variance $\sigma^2$ as $\hat{\sigma}^2 = \int \{\hat{G}_T(t,t) - \hat{K}_T(t,t)\} dt$. The fifth step is to use the algorithm described in Section 2.1 based on the estimates of the covariance functions $\hat{K}_T(t,s)$ and $\hat{K}_B(t,s)$.

**3. Principal component scores.** Estimating the principal component scores (PC scores) is straightforward in standard FPCA using, for example, direct numerical integration (henceforth denoted by NI). This can be done by plugging the estimators of $\mu(t)$ and $\phi_k(t)$ in the formula $\xi_k = \int_0^1 \{X(t) - \mu(t)\}\phi_k(t) dt$ and evaluating the integral over a grid of points. When measurement error is present, Yao et al. (2003) proposed shrinkage estimators of PC scores (henceforth denoted by NI-S). For sparse longitudinal data, Yao, Müller and Wang (2005) proposed the conditional expectation estimators PC scores (henceforth denoted by CE), which are also the best linear predictions based on observed data.

Estimating the scores in multilevel functional data is more complicated because the two sets of functional bases, $\{\phi_k^{(1)}(t)\}$ and $\{\phi_l^{(2)}(t)\}$, are not mutually orthogonal. We now describe our approach to estimate the scores,

MULTILEVEL FUNCTIONAL PRINCIPAL COMPONENT ANALYSIS 13

both with and without measurement error. In this section we assume that the PC scores $\xi_{ik}$ and $\zeta_{ijl}$ follow Gaussian distributions. We also discuss the consequences of the violation of these assumptions.

3.1. *Method 1: The full model.* For practical purposes, the infinite sums in the KL expansions are approximated by finite sums. Let $N_1, N_2$ be the number of dimensions that we decide to keep at levels 1 and 2, respectively. Once the fixed functional effects $\mu(t)$, $\eta_j(t)$, the eigenvalues $\lambda_k^{(1)}$, $\lambda_l^{(2)}$, and the eigenfunctions $\phi_k^{(1)}(t)$, $\phi_l^{(2)}(t)$ are estimated, the MFPCA model can be re-written as

$$(3.1) \quad \begin{cases} Y_{ij}(t) = \mu(t) + \eta_j(t) + \sum_{k=1}^{N_1} \xi_{ik}\phi_k^{(1)}(t) + \sum_{l=1}^{N_2} \zeta_{ijl}\phi_l^{(2)}(t) + \varepsilon_{ij}(t); \\ \xi_{ik} \sim N\{0, \lambda_k^{(1)}\}; \qquad \zeta_{ijl} \sim N\{0, \lambda_l^{(2)}\}; \qquad \varepsilon_{ij}(t) \sim N(0, \sigma^2), \end{cases}$$

where $\varepsilon_{ij}(t)$ appears only when functional data are observed with error. The Gaussian distributional assumption for the PC scores are for convenience, and we would discuss possible consequences of mis-specification. We will refer to this model as the full model for the PC scores (henceforth denoted by PC-F). A closer inspection of the model will reveal that this is a linear mixed model [Laird and Ware (1982)] with the random effects $\xi_{ik}$ and $\zeta_{ijl}$ being the quantities that we are trying to estimate. Thus, the mixed model inferential machinery can be used to estimate the scores using one of the two prototypical methods for estimating the random effects: (1) best linear unbiased prediction (BLUP) or (2) simulation of the posterior distribution using Markov Chain Monte Carlo (MCMC).

In this manuscript we choose to use MCMC, which is proved to work well, especially in scenarios with a large number of random effects, but the BLUP approach could be used as well. To specifically describe the procedure, we treat the estimates of $\mu(t)$, $\eta_j(t)$, $\lambda_k^{(1)}$ $\lambda_l^{(2)}$, $\phi_k^{(1)}(t)$, and $\phi_l^{(2)}(t)$, as fixed in model (3.1) (as obtained in the previous section). The prior for the precision parameter, $1/\sigma^2$, is assumed to be gamma with mean equal to 1 and a large variance (upon which we perform sensitivity analyses). The Markov chains would provide not only the point estimates (posterior means) but also the full posterior distribution for the principal component scores. More details on model specification and full conditionals can be found in the supplementary article [Di et al. (2009)].

This model is the generalization of the conditional expectation (CE) method proposed in Yao, Müller and Wang (2005) for single level functional data. It is appropriate to use for either dense or sparse functions. Using model (3.1) is appealing from a methodological perspective, but may raise computational challenges. Indeed, if $I$ subjects are observed at $J$ visits and

subject- and visit-specific functions are recorded on a grid of $T$ points, then model (3.1) has $IJT$ observations and $I(N_1 + JN_2)$ random effects. In our SHHS application in Section 5 the number of observations, $IJT$, exceeds 2.5 million, which may have computational consequences. In the following we present a solution that effectively circumvents the computational problems introduced by the volume of data.

3.2. *Method 2: The projection model.* In this section we present a simplified projection model related to model (3.1). The intuition behind the idea is to project each mean centered function into the space spanned by the eigenfunctions. Formally, we start by calculating

$$(3.2) \quad A_{ijk} := \int_0^1 \{Y_{ij}(t) - \mu(t) - \eta_j(t)\}\phi_k^{(1)}(t)\,dt$$

$$= \xi_{ik} + \sum_{l=1}^{N_2} \zeta_{ijl} c_{kl} + \epsilon_{ijk}^{(1)},$$

$$(3.3) \quad B_{ijl} := \int_0^1 \{Y_{ij}(t) - \mu(t) - \eta_j(t)\}\phi_l^{(2)}(t)\,dt$$

$$= \zeta_{ijl} + \sum_{k=1}^{N_1} \xi_{ik} c_{kl} + \epsilon_{ijl}^{(2)},$$

where $c_{kl} = \int_0^1 \phi_k^{(1)}(t)\phi_l^{(2)}(t)\,dt$ is the inner product of two eigenfunctions at different levels. Note that the residuals $\epsilon_{ijk}^{(1)}$ and $\epsilon_{ijl}^{(2)}$ incorporate both the measurement error as well as the residual variance in the discarded dimensions. For example, $\epsilon_{ijk}^{(1)} = \int_t \{\sum_{l=N_2+1}^{\infty} \zeta_{ijl}\phi_l^{(2)}(t) + \varepsilon_{ij}(t)\}\phi_k^{(1)}(t)\,dt$. The variance and covariance matrix of the integrated residuals are calculated in the supplementary article [Di et al. (2009)]. These indicate that the latter part would be dominant as long as $N_1$ and $N_2$ are large enough. Both $A_{ijk}$ and $B_{ijl}$ can be estimated by numerical integration from equations (3.2) and (3.3) by plugging in estimators of the corresponding eigenfunctions.

We rewrite equations (3.2) and (3.3) in matrix format. Let $\mathbf{A}_{ij} = (A_{ij1}, A_{ij2}, \ldots, A_{ijN_1})^T$, $\mathbf{B}_{ij} = (B_{ij1}, B_{ij2}, \ldots, B_{ijN_2})^T$, $\xi_i = (\xi_{i1}, \xi_{i2}, \ldots, \xi_{iN_1})^T$, $\zeta_{ij} = (\zeta_{ij1}, \zeta_{ij2}, \ldots, \zeta_{ijN_2})^T$, $\epsilon_{ij}^{(1)} = \{\epsilon_{ij1}^{(1)}, \epsilon_{ij2}^{(1)}, \ldots, \epsilon_{ijN_1}^{(1)}\}^T$, and $\epsilon_{ij}^{(2)} = \{\epsilon_{ij1}^{(2)}, \epsilon_{ij2}^{(2)}, \ldots, \epsilon_{ijN_2}^{(2)}\}^T$. Thus, equations (3.2) and (3.3) become

$$(3.4) \quad \begin{cases} \mathbf{A}_{ij} = \xi_i + C\zeta_{ij} + \epsilon_{ij}^{(1)}, \\ \mathbf{B}_{ij} = \zeta_{ij} + C^T \xi_i + \epsilon_{ij}^{(2)}, \\ \xi_i \sim N\{0, \Lambda^{(1)}\}, \quad \zeta_{ij} \sim N\{0, \Lambda^{(2)}\}, \\ \epsilon_{ij}^{(1)} \sim N(0, \sigma_1^2 \mathbf{I}_{N_1}), \quad \epsilon_{ij}^{(2)} \sim N(0, \sigma_2^2 \mathbf{I}_{N_2}), \end{cases}$$

where $C = (c_{kl})_{kl}$ is an $N_1 \times N_2$ matrix, $\Lambda^{(1)} = \mathrm{diag}\{\lambda_1^{(1)}, \lambda_2^{(1)}, \ldots, \lambda_{N_1}^{(1)}\}$, and

$\Lambda^{(2)} = \text{diag}\{\lambda_1^{(2)}, \lambda_2^{(2)}, \ldots, \lambda_{N_2}^{(2)}\}$. This is another linear mixed effects models with the residual variances $\sigma_1^2 = \text{var}\{\epsilon_{ijk}^{(1)}\}$ and $\sigma_2^2 = \text{var}\{\epsilon_{ijl}^{(2)}\}$ being estimated from the data. A heuristic justification for the independent and common variance assumption of $\epsilon_{ij}^{(1)}$ and $\epsilon_{ij}^{(2)}$ can be found in the supplementary article [Di et al. (2009)]. For future reference we call model (3.4) the projection model and denote it by PC-P. This model could be viewed as a generalization of the NI-S method [Yao et al. (2003)] single level setting, but with substantial complications due to the multilevel nature of our problem.

We use Bayesian framework implemented via MCMC for estimation of the principal component scores. As to the prior distribution for $1/\sigma_1^2$ and $1/\sigma_2^2$, we specify gamma priors with mean equal to 1 and very large variances. Estimation and inference on the principal component scores can be carried out in a similar way to the full model.

3.3. *Connection with the single level case.* The PC-P method is less computationally intensive than PC-F, because it summarizes each individual function, $Y_{ij}(t)$, using the low dimensional vectors $A_{ij}$ and $B_{ij}$. For dense functional data, the two methods yield similar results, and the PC-P method might be preferred. However, for sparse functional data, the PC-F method will typically perform better. To better understand our proposed methods, we compare them with methods for single level data. Table 1 provides a summary of applicability of methods to sparse and dense functional data. For example, the PC-P method, which is the multilevel counterpart of the NI-S method, works well only for dense functional data. The PC-F, which is the multilevel counterpart of the CE method, works well in general, but may be slow and numerically unstable for dense functional data.

One should not oversimplify the parallel between single and multilevel functional inferences. For example, in the case of dense functional data without noise, NI provides precise estimation of the true scores. In contrast, in the multilevel case NI estimates the scores with uncertainty. Indeed, consider the extreme case when the level 1 eigenfunction $\phi_1^{(1)}(t)$ and level 2 eigenfunction $\phi_1^{(2)}(t)$ are identical. The projection of the centered function $Y_{ij}(t) - \mu(t) - \eta_j(t)$ on it would estimate the sum of scores $\xi_{i1}$ and $\zeta_{ij1}$, but not the individual signals. Typically, estimation quality will depend on the inner products (matrix $C$) between level 1 and level 2 eigenfunctions. We illustrate this in simulation studies in Section 4.

TABLE 1
*Comparison of methods to estimate PC scores*

| Data structure | Method | Dense | | Sparse | |
| --- | --- | --- | --- | --- | --- |
| | | no noise | noise | no noise | noise |
| Single Level | NI | ✓ | | | |
| | NI-S | ✓ | ✓ | | |
| | CE | ✓ | ✓ | ✓ | ✓ |
| Multilevel | PC-P | ✓ | ✓ | | |
| | PC-F | ✓ | ✓ | ✓ | ✓ |

The two approaches via the linear mixed model framework are intrinsically parametric if we consider the numbers of dimensions $N_1$ and $N_2$ fixed. When the numbers of principal components are allowed to increase with sample size in an appropriate manner, the methods would be considered as nonparametric and targeting at the true infinite dimensional process.

The Gaussian assumptions in models (3.1) and (3.4) can be relaxed. The BLUPs of random effects have nice statistical properties that are robust to a wide range of departures from normality. They are best predictions under the normality assumption, and are best linear predictions in general.

**4. Simulation studies.** We evaluate the proposed methodology with extensive simulations. Throughout this section we generate samples of functions from the following model:

$$(4.1) \qquad Y_{ij}(t_m) = \sum_{k=1}^{4} \xi_{ik}\phi_k^{(1)}(t_m) + \sum_{l=1}^{4} \zeta_{ijl}\phi_l^{(2)}(t_m) + \varepsilon_{ij}(t_m),$$

where $\xi_{ik} \sim N(0, \lambda_k^{(1)})$, $\zeta_{ijl} \sim N(0, \lambda_l^{(2)})$, $\varepsilon_{ij}(t_m) \sim N(0, \sigma^2)$ and $\{t_m = \frac{m}{100} : m = 0, 1, \ldots, 100\}$.

We assume that there are $I = 200$ subjects (clusters), $J = 2$ visits per subject (measurements per cluster), $N_1 = 4$ eigenvalues and eigenfunctions at level 1 (subject), and $N_2 = 4$ eigenvalues and eigenfunctions at level 2 (visit). We use 200 subjects for illustration purposes, but our method applies to much larger data sets, such as the SHHS data. The true eigenvalues are $\lambda_k^{(1)} = 0.5^{k-1}, k = 1, 2, 3, 4$, and $\lambda_l^{(2)} = 0.5^{l-1}, l = 1, 2, 3, 4$, while $\mu(t) = 0$ and $\eta_j(t) = 0$. We consider a variety of scenarios, according to the choice of eigenfunctions and magnitude of noise, $\sigma = 0$ (no noise), $\sigma = 1$ (moderate), and $\sigma = 2$ (large). We conducted 1000 simulations for each scenario. This section provides details and results for two cases corresponding to the following choices of eigenfunctions:

*Case* 1. Mutually orthogonal bases.
  Level 1: $\phi_k^{(1)}(t) = \{\sqrt{2}\sin(2\pi t),\ \sqrt{2}\cos(2\pi t),\ \sqrt{2}\sin(4\pi t),\ \sqrt{2}\cos(4\pi t)\}$.
  Level 2: $\phi_l^{(2)}(t) = \{\sqrt{2}\sin(6\pi t),\ \sqrt{2}\cos(6\pi t),\ \sqrt{2}\sin(8\pi t),\ \sqrt{2}\cos(8\pi t)\}$.
*Case* 2. Mutually nonorthogonal bases.
  Level 1: same as in Case 1.
  Level 2: $\phi_1^{(2)}(t) = 1,\ \phi_2^{(2)}(t) = \sqrt{3}(2t-1),\ \phi_3^{(2)}(t) = \sqrt{5}(6t^2 - 6t + 1),$
  $\phi_4^{(2)}(t) = \sqrt{7}(20t^3 - 30t^2 + 12t - 1).$

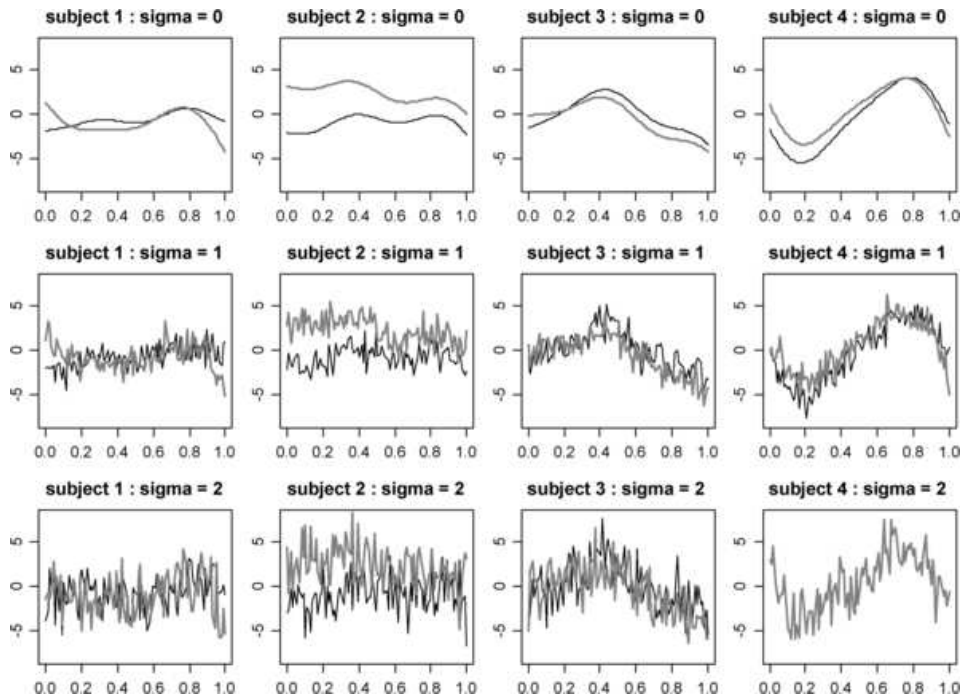

FIG. 2. *The simulated functions for four subjects: without and with noise. The curves in the first, second, and third row are functions with noise $\sigma = 0$, 1, and 2, respectively. In each figure, thin black and thick gray lines represent curves at visits 1 and 2, respectively.*

It can be verified that $\{\phi_k^{(1)} : k = 1, 2, 3, 4\}$ and $\{\phi_l^{(2)} : l = 1, 2, 3, 4\}$ are mutually orthogonal in Case 1, but not in Case 2. In the following, we focus on Case 2, because it is more realistic and relevant to our application to the SHHS in Section 5. The simulations for Case 1 are used to highlight the influence of the correlation between level 1 and level 2 functions on estimation quality for principal component scores. Results for Case 1 are provided in the supplementary article [Di et al. (2009)].

Figure 2 shows simulated functions for four subjects corresponding to different signal to noise ratios. As expected, when the amount of noise increases, the patterns at the subject level become less obvious or hardly recognizable. As we will show, our proposed methodology can recover the true signals with 200 samples even in the case when $\sigma = 2$.

4.1. *Eigenvalues and eigenfunctions.* Figure 3 shows estimated level 1 and 2 eigenvalues for the different magnitude of noise using the unsmooth MFPCA algorithm described in Section 2.2. This algorithm does not account for potential measurement error in the functional signal. The red solid line indicates the true eigenvalue. In the case of no noise ($\sigma = 0$), the eigenvalues can be recovered without bias. In the case of moderate noise ($\sigma = 1$), some small bias exists, especially for the third and fourth components. For large noise ($\sigma = 2$), this bias is more pronounced. These results have motivated us to develop the smooth MFPCA algorithm described in Section 2.3. Figure 4 shows similar results as in Figure 3 for the smooth MFPCA estimation algorithm, illustrating that the bias is practically removed.

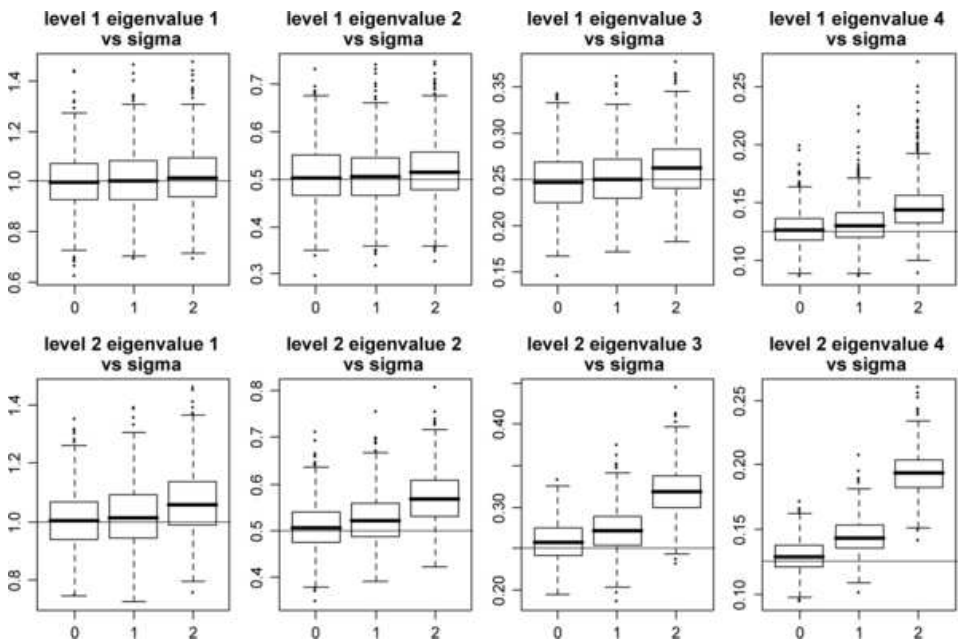

FIG. 3. *Boxplots of estimated eigenvalues using MFPCA: the true functions are without and with noise. The solid gray lines are the true eigenvalues.*

We now turn to estimating level 1 and 2 eigenfunctions. Figure 5 displays estimated eigenfunctions from 20 randomly selected simulations. Here the simulated data had no noise and we used the unsmooth version of our algorithm. Results indicate that the estimation method successfully separates level 1 and 2 variation and correctly captures the shape of each individual eigenfunction. When we increased the amount of noise, it is remarkable that both the unsmooth and smooth methods capture the overall shape of the eigenfunctions, even when the amount of noise is large. Moreover, the smooth MFPCA algorithm provides smoother curves, with each individual curve approximating well the true shape of the function. To save space, the corresponding figure is shown in the supplementary article [Di et al. (2009)].

4.2. *Principal component scores.* PC scores are central to our analyses, because they are the signals that will be used to assess the effect of sleep on health outcomes in subsequent analyses. In Section 3 we proposed two models, PC-F and PC-P, to estimate PC scores for multilevel functional principal component analysis. Here we compare the performance for these two models with respect to the root mean square errors (RMSE). In each scenario, we randomly selected 10 simulated datasets, and estimated principal component scores using posterior means from the Markov chains. We ran several chains, with different initial values, and verified that they had good convergence and mixing properties. More details can be found in the supplementary article [Di et al. (2009)] for this paper. We found that the root mean square errors are very stable across simulated datasets. Thus, we report RMSEs based on 10 simulated datasets.

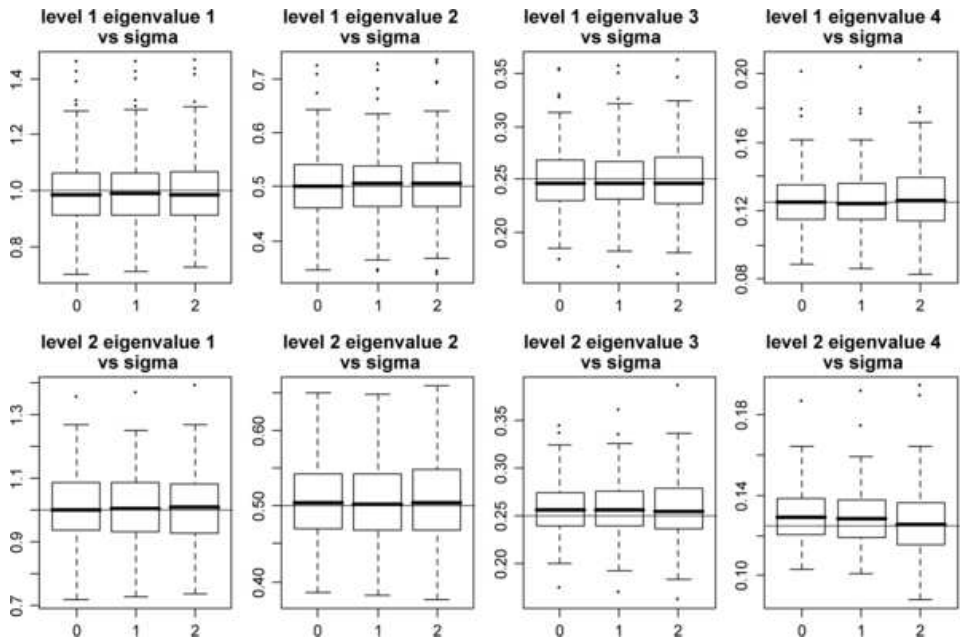

Fig. 4. *Boxplots of estimated eigenvalues using the smooth MFPCA: the true functions are without and with noise. The solid gray lines are the true eigenvalues.*

Table 2 summarizes results for several scenarios. In Case 1, when level 1 and 2 eigenfunctions are mutually orthogonal, the estimation of scores depends only on the estimation quality of the covariance matrices. Score estimates approximate the true scores well, as indicated by the smaller RMSE compared to those obtained in Case 2. For either PC-F or PC-P, the RMSE is smaller when there is no noise ($\sigma = 0$), and larger for noisy data ($\sigma = 2$). Even though PC-F performed slightly better than PC-P, the latter might still be a nice practical choice, especially considering computational time.

In Case 2, similar findings are observed. In addition, the RMSEs are generally larger in Case 2 at the same level of residual noise. This is due to the nonorthogonality between eigenfunctions at the two levels, as discussed in Section 3. For example, the inner product of level 1 eigenfunction 2 and level 2 eigenfunction 3 is 0.96, corresponding to a 16 degree angle between them. Such an angle could be expected to affect estimation quality, which is confirmed by the relatively large RMSE for these two components.

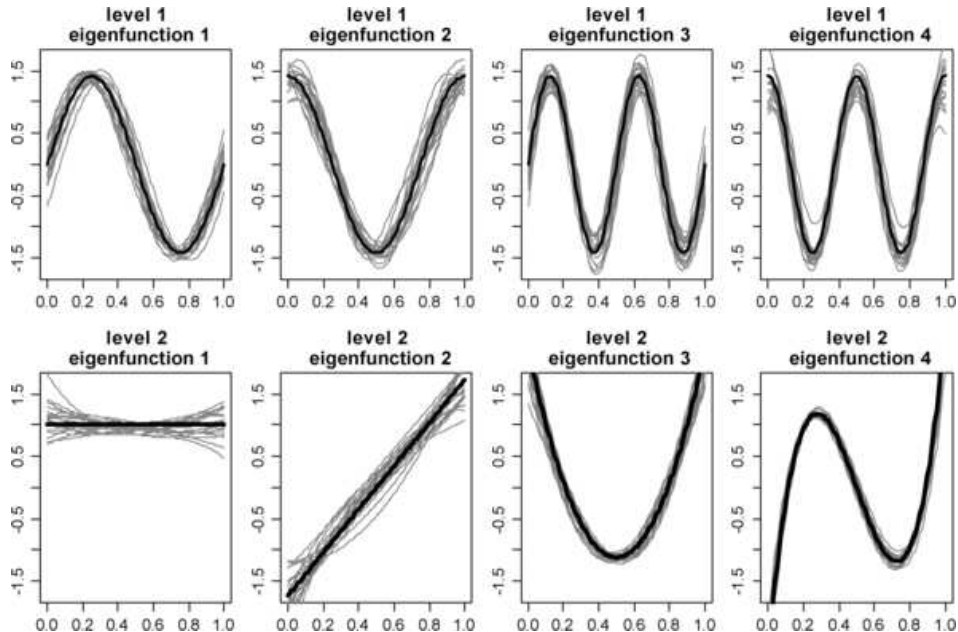

Fig. 5. *Estimated eigenfunctions when the functions are observed without noise, that is, $\sigma = 0$, from 20 randomly selected simulations. Thick black lines: true eigenfunctions.*

**5. The analysis of sleep data from the SHHS.** We now apply our proposed methods to the SHHS data. Two approaches are implemented and compared. The first approach is to smooth the raw data and use the smoothed functional estimates in the MFPCA framework described in Section 2.2. The second approach uses the raw (unsmoothed) data in the smooth MFPCA framework described in Section 2.3. Because results are similar, we only present results for the second approach.

For our analyses, we considered 3201 subjects with complete visit 1 and visit 2 data and sleep duration that exceeds 4 hours at both visits, and we analyzed data for the first 4 hours of sleep. Figure 6 displays the estimated overall and visit-specific normalized sleep EEG $\delta$-power functions. The left panel shows the raw estimates and the right panel shows estimated mean functions after smoothing each curve. The smaller mean function at visit 2 is most likely associated with the 5 year aging of the cohort.

5.1. *Eigenvalues and eigenfunctions.* For the subject level (or level 1), Table 3 displays the estimated eigenvalues. Interestingly, most of the subject level information is contained in only 2–3 dimensions. For example, the first eigenvalue explains 80.6% of the variation, while the second and third eigenvalues explain 7.7% and 3.7% of variation, respectively. Together, they explain more than 91% of the subject level variation. Figure 7 shows the first three subject level eigenfunctions. The bottom panels display the population mean function $\mu(t)$ and the functions obtained by adding and subtracting a suitable multiple of the eigenfunctions to the mean, that is, $\mu(t) + \sqrt{\lambda_k^{(1)}} \cdot \phi_k^{(1)}(t)$, and $\mu(t) - \sqrt{\lambda_k^{(1)}} \cdot \phi_k^{(1)}(t)$. Plus signs indicate addition and minus signs subtraction. Such plots are helpful to understand the variability in the direction of certain eigenfunctions. The first eigenfunction is positive, indicating that subjects with positive scores on this component will tend to get a consistently larger proportion of sleep EEG $\delta$-power than the population average. The second eigenfunction displays an oscillatory component. Subjects with positive scores will have less sleep EEG $\delta$-power in the first 2 hours and slightly more between hours 2 and 4. These components are easy to interpret scientifically, but they would be difficult to identify and quantify by direct inspection of subject plots.

Visit level (or level 2) has more directions of variation. Indeed, 90% of the variability is explained by the first 14 principal components, with 50% of the variability being explained by the first 4 components. The spread of variability over many components is not surprising for such a large data set and reflects the large within-subject heterogeneity. Figure 8 shows the first 3 estimated principal components at level 2. For interpretation, it is helpful to remember that level 2 eigenfunctions represent the random visit-specific functional deviation from the subject specific function. For example, the first principal component is positive, indicating that subjects who are loaded on this component are more prone to a shift in average EEG δ-power, that is, subject visits with a positive score on this component correspond to a larger proportion of $\delta$ sleep than the average. The difference is more pronounced after the first hour. Eigenfunctions beyond the first are typically periodic.

TABLE 2
*Root mean square errors for estimating scores using methods PC-F and PC-P*

| Method | $\sigma$ | Level 1 component | | | | Level 2 component | | | |
|---|---|---|---|---|---|---|---|---|---|
| | | 1 | 2 | 3 | 4 | 1 | 2 | 3 | 4 |
| Case 1: PC-F | 0 | 0.097 | 0.146 | 0.072 | 0.047 | 0.122 | 0.143 | 0.124 | 0.093 |
| | 2 | 0.199 | 0.207 | 0.144 | 0.140 | 0.221 | 0.222 | 0.236 | 0.213 |
| Case 1: PC-P | 0 | 0.112 | 0.155 | 0.082 | 0.049 | 0.127 | 0.154 | 0.127 | 0.095 |
| | 2 | 0.206 | 0.212 | 0.145 | 0.139 | 0.222 | 0.229 | 0.237 | 0.217 |
| Case 2: PC-F | 0 | 0.196 | 0.202 | 0.114 | 0.080 | 0.139 | 0.152 | 0.128 | 0.105 |
| | 2 | 0.415 | 0.385 | 0.174 | 0.153 | 0.246 | 0.347 | 0.368 | 0.263 |
| Case 2: PC-P | 0 | 0.210 | 0.207 | 0.120 | 0.080 | 0.142 | 0.167 | 0.129 | 0.112 |
| | 2 | 0.410 | 0.408 | 0.182 | 0.160 | 0.252 | 0.355 | 0.393 | 0.273 |

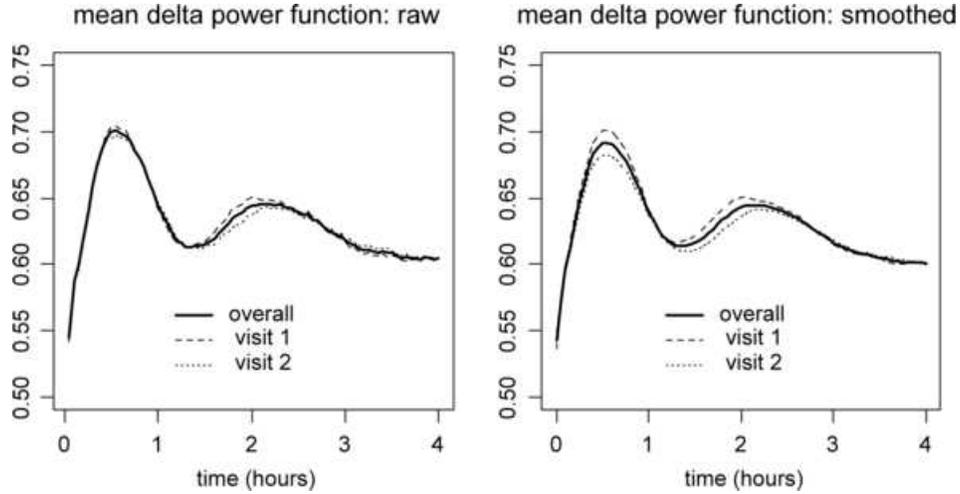

Fig. 6. *Estimated overall and visit-specific mean δ-power functions for the SHHS data. The left panel shows the raw mean functions, while the right panel shows the mean functions after smoothing each curve.*

Table 3
*Estimated eigenvalues on both levels for SHHS data using MFPCA. Three components are kept for level 1 (subject level), and 14 components are kept for level 2. "percent var" stands for the percentage of variance explained by the component, and "cum percent var" means the cumulative percentage of variance explained*

| Level 1 eigenvalues | | | | | | | |
|---|---|---|---|---|---|---|---|
| Component | 1 | 2 | 3 | | | | |
| eigenvalue ($\times 10^{-3}$) | 13.00 | 1.24 | 0.55 | | | | |
| percent var | 80.59 | 7.68 | 3.38 | | | | |
| cum percent var | 80.59 | 88.27 | 91.66 | | | | |
| Level 2 eigenvalues | | | | | | | |
| Component | 1 | 2 | 3 | 4 | 5 | 6 | 7 |
| eigenvalue ($\times 10^{-3}$) | 12.98 | 7.60 | 7.46 | 6.45 | 5.70 | 4.47 | 3.07 |
| percent var | 21.84 | 12.79 | 12.55 | 10.85 | 9.58 | 7.52 | 5.17 |
| cum percent var | 21.84 | 34.63 | 47.17 | 58.02 | 67.61 | 75.13 | 80.30 |
| Component | 8 | 9 | 10 | 11 | 12 | 13 | 14 |
| eigenvalue ($\times 10^{-3}$) | 2.17 | 1.72 | 1.35 | 1.12 | 0.90 | 0.75 | 0.59 |
| percent var | 3.65 | 2.89 | 2.28 | 1.88 | 1.51 | 1.26 | 0.99 |
| cum percent var | 83.95 | 86.85 | 89.13 | 91.00 | 92.51 | 93.77 | 94.76 |

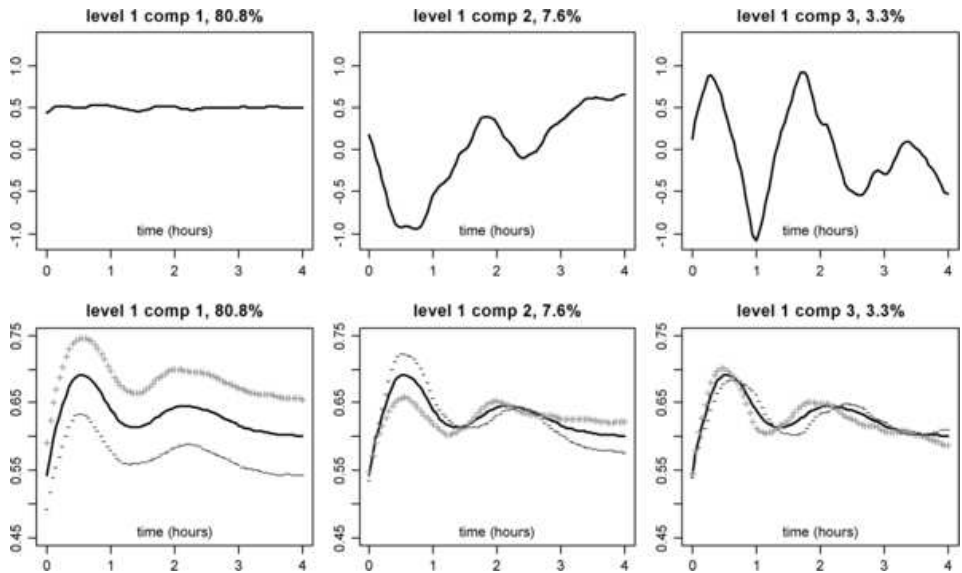

FIG. 7. *The first three level 1 (between subject level) eigenfunctions using MFPCA. The three upper panels show the estimated eigenfunctions versus time in hours. The bottom panels display the population mean function and the functions obtained by adding and subtracting a suitable multiple of the eigenfunctions to the mean. Plus signs indicate addition and minus signs indicate subtraction.*

The proportion of variability explained by subject level functional clustering, $\rho_W$, was defined in Section 2.2. In the SHHS we estimate $\hat{\rho}_W = 0.213$, that is, 21.3% of variability in the sleep EEG $\delta$-power is attributable to the subject level variability. Another interpretation is that the average correlation between two functions from the same subject is 0.213.

Because within-subject correlation is small, one may wonder whether it is due to random variation. To test the null hypothesis $H_0 : \rho_W = 0$ versus $H_A : \rho_W > 0$, we used a parametric bootstrap as follows. We fitted a model to the sleep data under $H_0$ and kept the first 14 level 2 eigenfunctions. Based on the estimated model, we generated bootstrap samples and extracted eigenvalues and eigenfunctions. Based on 1000 bootstrap samples using the thresholds described in Section 2.2, the 95% confidence interval for $\rho_W$ is $[0.011, 0.024]$, which does not contain 0.213, indicating that there is strong evidence of within-subject correlation, even though the two visits are 5 years apart. We also carried out a parametric bootstrap under the alternative hypothesis $H_1 : \rho_W \neq 0$. The data are simulated using 3 true level 1 and 14 level 2 eigenvalues and eigenfunctions estimated from SHHS. Under $H_1$. the 95% confidence interval for $\rho$ is $[0.210, 0.236]$, covering the estimated 0.213 for the sleep data.

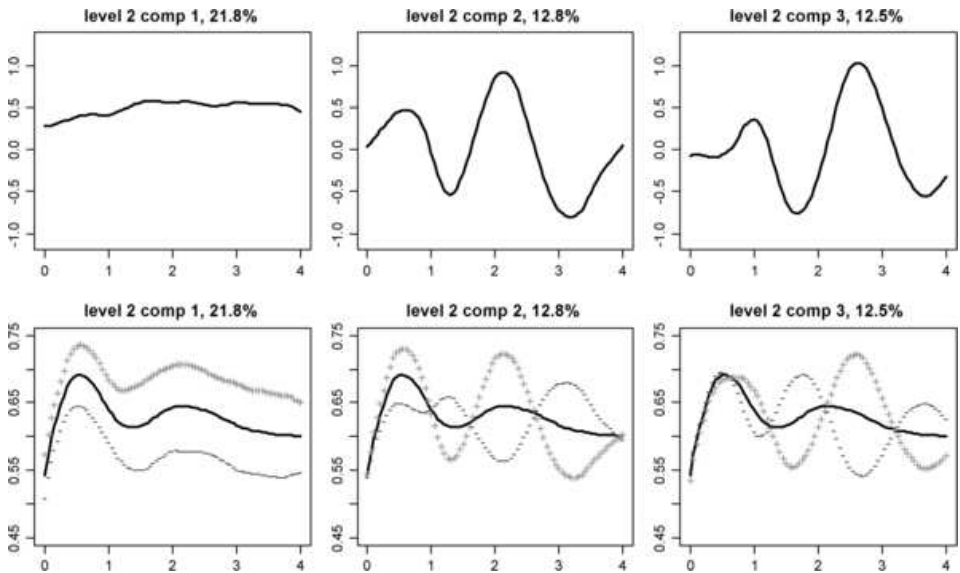

Fig. 8. *The first three level 2 (within subject level) eigenfunctions using MFPCA.*

5.2. *Distribution of PC scores.* One of the main goals of PCA is to provide dimensionality reduction. For example, the infinite dimensional subject-specific functions in SHHS have a representation in terms of 3-dimensional vectors of scores. This low-dimensional representation can than be used in subsequent analyses, by using the scores either as covariates or outcomes.

As discussed in Section 3, we estimated the principal component scores via a Bayesian extension of the model using MCMC. We chose inverse gamma priors with large variances for the variance components $\sigma^2$, $\sigma_1^2$, and $\sigma_2^2$. The Markov chains were monitored and diagnosed to have good convergence and mixing properties. We show results from the "projection" model (PC-P), since it is less computationally intensive and it is shown to perform well for dense functional data (Section 4). Figure 9 displays the distribution of the estimated subject-specific scores. The upper left panel shows the scatterplot of the first and second PC scores, indicating that the normality assumption is reasonable. The first component explains more than 80% of the variation and is, basically, a vertical shift. Subjects who have high scores on this component tend to have a higher percentage of delta power sleep. Note that the first component scores have a much wider range that the second component scores, which is consistent with the much larger amount of variability explained by the first component. The other panels show the distribution of the first PC scores versus covariates. Figure 9 indicates that females tend to have a higher average percent sleep EEG $\delta$-power sleep than males. Compared to nonsmokers, former smokers seems to have less percent sleep EEG $\delta$-power, while current smokers have an even smaller percentage sleep EEG $\delta$-power. The mean $\delta$-power sleep seems to decrease with age, and RDI, though the trend is more clear for age. BMI and the mean $\delta$-power do not exhibit a clear association.

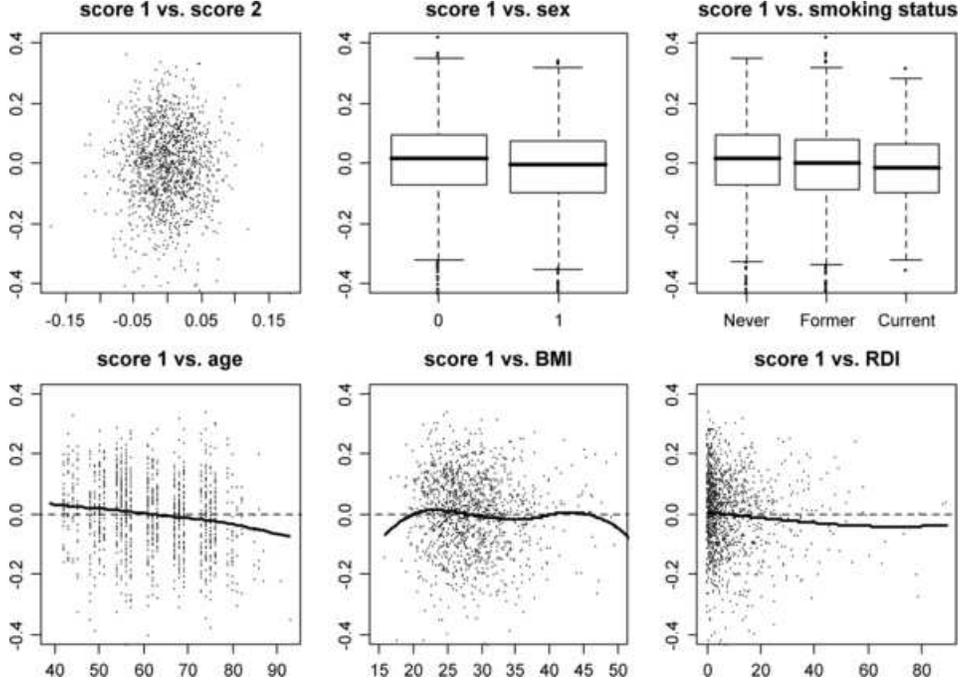

Fig. 9. *Estimated subject-specific principal component scores. Upper left panel: scatterplot of the* 1*st versus* 2*nd PC scores. Other panels: distribution of the first PC scores versus five covariates: sex, smoking status, age, BMI, and RDI. For the sex variable, female is the reference group.*

5.3. *Association between component scores and hypertension.* In this section we analyze the association between percent $\delta$-power in the sleep EEG and hypertension. Consider the functional generalized linear regression model

$$\text{(5.1)} \qquad \text{logit}\{P(Y_i = 1)\} = \beta_0 + \int \beta(t) Z_i(t)\, dt + V_i^T \gamma,$$

where $Y_i$ is the binary outcome indicating whether the subject has hypertension, $Z_i(t)$ is the subject level $\delta$-power function, and $V_i$ is a vector of other covariates (such as age, sex, etc). Since the subject level eigenfunction $\{\phi_k^{(1)}(t)\}$ is an orthogonal basis, $\beta(t)$ can be written as $\beta(t) = \sum_{k=1}^{N_1} \beta_k \phi_k^{(1)}(t)$. Thus, equation (5.1) becomes

$$\text{(5.2)} \qquad \text{logit}\{P(Y_i = 1)\} = \beta_0 + \sum_{k=1}^{N_1} \beta_k \xi_{ik} + V_i^T \gamma,$$

which is a logistic regression model with subject level PC scores as covariates.

We fitted several models with various levels of confounding adjustment, with results summarized in Table 4. Models included combinations of confounders including sex, smoking status (with three categories: never smokers, former smokers, and current smokers), age, body mass index (BMI), and respiratory disturbance index (RDI). All six models indicated that the first principal component score is strongly and negatively associated with hypertension. The magnitude of association varies with the amount of confounding adjustment. For example, Model 2 estimates that a subject with one unit higher in the first principal component has $e^{-1.59} = 0.204$ ($p$-value: $< 0.001$) times the odds of hypertension, controlling for sex and smoking status. Considering the scale of the PC scores (the first component PC scores have mean zero and standard deviation 0.11), standardized coefficients would be easier to interpret. After standardizing, one standard deviation increase in the first PC score is associated with an odds ratio $e^{-0.205} = 0.815$ ($p$-value: $< 0.001$). Model 6, which adjusts for all the confounders, estimated an odds ratio of $e^{-0.86} = 0.423$ per unit increase in the first PC score, or an odds ratio $e^{-0.11} = 0.895$ per one standard deviation increase in the first PC score. The second and third principal components were not found to be associated with hypertension.

The negative relationship between smoking and hypertension may seem counterintuitive. However, in this study smokers are younger, have a lower body mass index, and many other smokers with severe disease were not included in the study [Zhang et al. (2006)]. A simple exploratory analysis ignoring the sleep EEG measurements showed similar results.

Since the first principal component scores are approximately the shrinkage estimators of the mean $\delta$ power, one might propose to simply summarize the functional data by the subject level mean $\delta$ power without the principal component analysis. It is true that the mean $\delta$ power predicts hypertension. However, without our analyses, choosing the mean $\delta$ power would have been just a lucky heuristic guess. A researcher looking for predictors would probably not be satisfied with taking the average because of the potential loss of information. Our methods provide a systematic way of looking for predictors in a large functional space. We are especially interested in the other directions of variation. Our analyses show that the loss of information would be minimal, as far as prediction of hypertension in the SHHS is concerned. Things could be very different for another outcome and data set. As a final point, our methods provide shrinkage estimates of the principal component scores. In the second-stage functional regression (5.2), our shrinkage estimators lead to unbiased effect estimators, while empirical means lead to biased effect estimators. These points are further discussed in Section 6.

TABLE 4
*Models for association between hypertension and sleep EEG δ-power. Smoking status has three categories: never smokers (reference), former smokers (smk:former) and current smokers (smk:current). For the variable sex, female is the reference group and an asterisk indicates significance at level* 0.05

|  | Model 1 | Model 2 | Model 3 | Model 4 | Model 5 | Model 6 |
|---|---|---|---|---|---|---|
| Score 1 | −1.58 (0.28)* | −1.59 (0.28)* | −1.48 (0.29)* | −1.07 (0.30)* | −1.50 (0.28)* | −0.86 (0.30)* |
| Score 2 | 0.66 (0.97) | 0.74 (0.97) | 0.51 (0.99) | 0.14 (1.01) | 0.46 (0.98) | −0.26 (1.04) |
| Score 3 | 1.74 (1.56) | 1.66 (1.56) | 1.85 (1.59) | −0.28 (1.63) | 1.85 (1.57) | −0.26 (1.67) |
| Sex |  | 0.09 (0.07) | 0.11 (0.07) | 0.12 (0.08) | −0.00 (0.08) | 0.10 (0.08) |
| smk:former |  | −0.08 (0.08) | −0.11 (0.08) | −0.14 (0.08) | −0.08 (0.08) | −0.19 (0.08)* |
| smk:current |  | −0.28 (0.12)* | −0.30 (0.12)* | −0.13 (0.12) | −0.24 (0.12)* | −0.11 (0.13) |
| Age |  |  |  | 0.06 (0.00)* |  | 0.06 (0.00)* |
| BMI |  |  | 0.05 (0.01)* |  |  | 0.06 (0.01)* |
| RDI |  |  |  |  | 0.02 (0.00)* | 0.01 (0.00)* |

5.4. *Summary of SHHS results.* Though we defer a complete analysis of the SHHS data to companion manuscripts, we briefly summarize these results and how they fit into the broader context of sleep research. In this manuscript we have characterized the geometric directions of variation of slow-wave sleep patterns, both within- and between-subjects. The results of this manuscript represent the first study of these patterns. Moreover, the quality and comprehensiveness of the underlying dataset place further emphasis on the inter-subject results. The results show that the principal direction of variation is an overall shift in slow-wave sleep, and that this explains a large proportion of the geometric variation. The second component represents an early-night shift in slow-wave sleeping. These important findings will allow researchers to focus on simple metrics with the knowledge that they explain a majority of the functional variation.

The finding that the first principal component was associated with a lower odds ratio for hypertension is scientifically and clinically relevant. Perhaps most interestingly, the effect persists with the inclusion of the respiratory disturbance index in the model. That is, the first principal component appears to capture a component of vascular health that is independent of the most well-established link between sleep disruption and cardiac health. It is well-established that the $\delta$-power in the EEG, a characteristic of slow-wave sleep, reflects that homeostatic need for sleep [Borbely and Achermann (1999)]. It increases proportionally in relation to prior waking and decreases during sleep. With increasing age, the amount of slow wave activity (i.e., $\delta$-power) declines and is accompanied by significant decreases in growth hormone secretion and higher levels of circulating cortisol [Van Cauter, Leproult and Plat (2000)]. More recent data also indicate that slow wave sleep may also play a role in brain restoration and memory consolidation [Massimini et al. (2007)]. Collectively, the available data suggests that, in the absence of other intrinsic or extrinsic sleep disorders, higher $\delta$-power EEG during sleep may be associated with favorable health profiles. To date, there are no studies correlating the amount of $\delta$-power EEG during sleep with prevalent cardiovascular conditions. If the observed association between $\delta$-power EEG and prevalent hypertension is in fact causal, it would implicate poor sleep quality as an important determinant of chronic health conditions.

**6. Discussion.** The SHHS contains, by far, the largest EEG collection of sleep-related data on a community cohort at multiple visits. Important challenges raised by the SHHS data, but common to many other modern data sets, are the large dimensionality, the large subject-specific measurement error, and the large within- and between-subject variability. The powerful MFPCA methods developed in this paper allow a robust and computationally feasible decomposition of the observed functional variability. MFPCA provides a parsimonious geometric decomposition of subject level and visit/subject level functional characteristics, which would be impossible to detect by a simple inspection of the plots for three or three thousand subjects.

We appreciate questions, as posed by the Associate Editor and others, of whether multilevel analysis in necessary in this setting. For example, one might propose a simpler and more convenient approach, say, averaging the replicate functions per subject and conducting a single-level principal component analysis. However, we propose and largely demonstrate that a multilevel analysis is necessary to correctly identify and quantify the subject-specific and subject/visit-specific variability. For example, the MFPCA analysis of the SHHS data identifies 3 directions of variation for the long term subject-average, whereas a single-level FPCA would identify 16 or 17. Most of these directions of variation would be drowned in noise induced by the subject/visit-specific variation. To better understand this point, note that in the case of no measurement error, the covariance function for $\bar{X}_i.(t)$ is $K_B(s,t) + K_W(s,t)/J > K_B(s,t)$. Thus, the eigenfunctions extracted from the simple PCA approach do not capture the true subject level functional variability. In the SHHS application, it is even worse, because the number of visits is small, $J = 2$, and the eigenvalues of $K_W(s,t)$ are larger than those of $K_B(s,t)$.

Moreover, multi-level analysis provides the following: (1) a parsimonious decomposition of the hierarchical functional space; (2) shrinkage estimators of the principal component scores by borrowing strength across subjects; (3) subject-specific functional predictions and uncertainty of such predictions; and (4) measures of functional correlation. The shrinkage estimators of the principal component scores typically outperform the raw estimates in terms of mean square error. The principal component scores, which summarize the high dimensional functional data by multivariate vectors, are often used as predictors in the subsequent analysis. It can be shown that in a functional regression context our shrinkage estimators lead to unbiased effect estimators, while the raw estimators lead to biased regression coefficients. The comparison is analogous to that between the regression calibration estimates and the naive estimates in measurement error models [Carroll et al. (2006)].

One limitation of our methodology is that it does not preserve the infinite dimensionality of the functional space in a strict sense. Instead, it relies on the assumption that the finite dimensional functional spaces are well approximated by finite and small dimensional subspaces. Thus, once we condition on these functional subspaces, our approach becomes inherently parametric. However, when the number of principal components are allowed to increase with sample size in an appropriate manner, the method would be considered as nonparametric and targeting at the true process.

A major contribution of this paper is that we opened many theoretical and applied problems in an area of statistical research that is likely to have a significant impact on the analysis of increasingly complex data sets. However, much research remains to be done in this area. First, it would be important to improve the covariance function estimators and ensure that they are positive-definite and efficient. Second, estimating the dimension of the functional spaces is ultimately equivalent to a sequential battery of tests for zero variance in linear mixed effects. Third, plugging in the BLUP or posterior means of scores in second level analyses and ignoring their associated variability may lead to biased results in nonlinear second level analyses.

## SUPPLEMENTARY MATERIAL

**Multilevel functional principal component analysis**
(DOI: [10.1214/08-AOAS206SUPP](10.1214/08-AOAS206SUPP); .pdf). We assess the criterion for choosing the number of principal components, provide details for Bayesian MCMC for estimating principal component scores, and show additional results for simulations and the application to SHHS. We also provide some technical details for the variance and covariance of the residuals from the projection model.

## REFERENCES

- BALADANDAYUTHAPANI, V., MALLICK, B. K., HONG, M. Y., LUPTON, J. R., TURNER, N. D. and CARROLL, R. J. (2008). Bayesian hierarchical spatially correlated functional data analysis with application to colon carcinogenesis. *Biometrics* **64** 64–73.
- BESSE, P. and RAMSAY, J. O. (1986). Principal components analysis of sampled functions. *Psychometrika* **51** 285–311. [MR0848110](MR0848110)
- BIGELOW, J. L. and DUNSON, D. B. (2007). Bayesian adaptive regression splines for hierarchical data. *Biometrics* **63** 724–732. [MR2395709](MR2395709)
- BORBELY, A. A. and ACHERMANN, P. (1999). Sleep homeostasis and models of sleep regulation. *J. Biological Rhythms* **14** 557.
- BRUMBACK, B. A. and RICE, J. A. (1998). Smoothing spline models for the analysis of nested and crossed samples of curves. *J. Amer. Statist. Assoc.* **93** 961–976. [MR1649194](MR1649194)
- CARRIER, J., LAND, S., BUYSSE, D. J., KUPFER, D. J. and MONK, T. H. (2001). The effects of age and gender on sleep EEG power spectral density in the middle years of life (ages 20–60 years old). *Psychophysiology* **38** 232–242.
- CARROLL, R. J., RUPPERT, D., STEFANSKI, L. A. and CRAINICEANU, C. M. (2006). *Measurement Error in Nonlinear Models: A Modern Perspective*, 2nd ed. Chapman and Hall/CRC Press, Boca Raton, FL. [MR2243417](MR2243417)
- CRAINICEANU, C. M., CAFFO, B. S., DI, C.-Z. and NARESH, P. M. (2009). Nonparametric signal extraction and measurement error in the analysis of electroencephalographic activity during sleep. *J. Amer. Statist. Assoc.* To appear.

Di, C.-Z., Crainiceanu, C. M., Caffo, B. S. and Punjabi, N. M. (2009). Supplment to "Multilevel functional principal component analysis." DOI: 10.1214/08-AOAS206SUPP.

Diggle, P. J., Heagerty, P., Liang, K.-Y. and Zeger, S. L. (2002). *The Analysis of Longitudinal Data*. Oxford Univ. Press. MR2049007

Fan, J. and Gijbels, I. (1996). *Local Polynomial Modelling and Its Applications. Monographs on Statistics and Applied Probability* **66**. Chapman & Hall, London. MR1383587

Goldstein, H. (1995). *Multilevel Statistical Models*. A Hodder Arnold Publication, London.

Guo, W. (2002). Functional mixed effects models. *Biometrics* **58** 121–128. MR1891050

Hall, P. and Hosseini-Nasab, M. (2006). On properties of functional principal components analysis. *J. R. Stat. Soc. Ser. B Statist. Methodol.* **68** 109–126. MR2212577

Hall, P., Müller, H.-G. and Yao, F. (2008). Modeling sparse generalized longitudinal observations with latent Gaussian processes. *J. R. Stat. Soc. Ser. B Statist. Methodol.* **70** 703–723.

Indritz, J. (1963). *Methods in Analysis*. Macmillan, New York. MR0150991

James, G. M., Hastie, T. J. and Sugar, C. A. (2000). Principal component models for sparse functional data. *Biometrika* **87** 587–602. MR1789811

Karhunen, K. (1947). *Über lineare Methoden in der Wahrscheinlichkeitsrechnung*. Suomalainen Tiedeakatemia. MR0023013

Laird, N. and Ware, J. (1982). Random-effects models for longitudinal data. *Biometrics* **38** 963–974.

Lin, X. and Carroll, R. J. (2000). Nonparametric function estimation for clustered data when the predictor is measured without/with error. *J. Amer. Statist. Assoc.* **95** 520–534. MR1803170

Loève, M. (1945). Fonctions aléatoires de second ordre. *C. R. Acad. Sci.* **220** 469.

Massimini, M., Ferrarelli, F., Esser, S. K., Riedner, B. A., Huber, R., Murphy, M., Peterson, M. J. and Tononi, G. (2007). Triggering sleep slow waves by transcranial magnetic stimulation. *Proc. Natl. Acad. Sci.* **104** 84–96.

Morris, J. S., Brown, P. J., Herrick, R. C., Baggerly, K. A. and Coombes, K. R. (2008). Bayesian analysis of mass spectrometry proteomic data using wavelet-based functional mixed models. *Biometrics* **64** 479–489.

Morris, J. S. and Carroll, R. J. (2006). Wavelet-based functional mixed models. *J. Roy. Statist. Soc. Ser. B* **68** 179–199. MR2188981

Morris, J. S., Vannucci, M., Brown, P. J. and Carroll, R. J. (2003). Wavelet-based nonparametric modeling of hierarchical functions in colon carcinogenesis. *J. Amer. Statist. Assoc.* **98** 573–584. MR2011673

Müller, H.-G. (2005). Functional modelling and classification of longitudinal data. *Scand. J. Statist.* **32** 223–240. MR2188671

Quan, S. F., Howard, B. V., Iber, C., Kiley, J. P., Nieto, F. J., O'Connor, G. T., Rapoport, D. M., Redline, S., Robbins, J., Samet, J. M. and Wahl, P. W. (1997). The sleep heart health study: Design, rationale, and methods. *Sleep* **20** 1077–1085.

Ramsay, J. O. and Dalzell, C. J. (1991). Some tools for functional data analysis. *J. R. Stat. Soc. Ser. B Statist. Methodol.* **53** 539–572. MR1125714

Ramsay, J. O. and Silverman, B. W. (2005). *Functional Data Analysis*. Springer. MR2168993

Raudenbush, S. W. and Bryk, A. S (2002). *Hierarchical Linear Models: Applications and Data Analysis Methods*, 2nd ed. *Advanced Quantitative Techniques in the Social Sciences Series* **1**. Sage, Thousand Oaks, CA.

Rice, J. A. and Silverman, B. W. (1991). Estimating the mean and covariance structure nonparametrically when the data are curves. *J. R. Stat. Soc. Ser. B Statist. Methodol.* **53** 233–243. MR1094283

Ruppert, D., Wand, M. P. and Carroll, R. J. (2003). *Semiparametric Regression. Cambridge Series in Statistical and Probabilistic Mathematics* **12**. Cambridge Univ. Press, Cambridge. MR1998720

Shahar, E., Whitney, C. W., Redline, S., Lee, E. T., Newman, A. B., Javier Nieto, F., O'connor, G. T., Boland, L. L., Schwartz, J. E. and Samet, J. M. (2001). Sleep-disordered breathing and cardiovascular disease cross-sectional results of the Sleep Heart Health Study. *American J. Respiratory and Critical Care Medicine* **163** 19–25.

Silverman, B. W. (1996). Smoothed functional principal components analysis by choice of norm. *Ann. Statist.* **24** 1–24. MR1389877

Sing, H. C., Kautz, M. A., Thorne, D. R., Hall, S. W., Redmond, D. P., Johnson, D. E., Warren, K., Bailey, J. and Russo, M. B. (2005). High-frequency EEG as measure of cognitive function capacity: A preliminary report. *Aviation, Space and Environmental Medicine* **76** C114–C135.

Staniswalis, J. G. and Lee, J. J. (1998). Nonparametric regression analysis of longitudinal data. *J. Amer. Statist. Assoc.* **93** 1403–1404. MR1666636

Tassi, P., Bonnefond, A., Engasser, O., Hoeft, A., Eschenlauer, R. and Muzet, A. (2006). EEG spectral power and cognitive performance during sleep inertia: the effect of normal sleep duration and partial sleep deprivation. *Physiological Behavior* **87** 177–184.

Van Cauter, E., Leproult, R. and Plat, L. (2000). Age-related changes in slow wave sleep and rem sleep and relationship with growth hormone and cortisol levels in healthy men. *J. Amer. Medical Assoc.* **284** 861–868.

Yao, F., Müller, H.-G., Clifford, A. J., Dueker, S. R., Follett, J., Lin, Y., Buchholz, B. A. and Vogel, J. S. (2003). Shrinkage estimation for functional principal component scores with application to the population kinetics of plasma folate. *Biometrics* **59** 676–685. MR2004273

Yao, F., Müller, H.-G. and Wang, J. L. (2005). Functional data analysis for sparse longitudinal data. *J. Amer. Statist. Assoc.* **100** 577–591. MR2160561

Zhang, L., Samet, J., Caffo, B. S. and Punjabi, N. M. (2006). Cigarette smoking and nocturnal sleep architecture. *American Journal of Epidemiology* **164** 529.

C.-Z. Di
C. M. Crainiceanu
B. S. Caffo
Department of Biostatistics
Johns Hopkins University
615 N. Wolfe Street
Baltimore, Maryland 21205
USA
E-mail: cdi@jhsph.edu
ccrainic@jhsph.edu
bcaffo@jhsph.edu
URL: http://www.biostat.jhsph.edu

N. M. Punjabi
Department of Epidemiology
Johns Hopkins University
615 N. Wolfe Street
Baltimore, Maryland 21205
USA